\documentclass[fleqn,twoside]{article}
\usepackage{amsmath}
\usepackage{espcrc2}
\usepackage{amssymb}
\usepackage{psfrag}
\usepackage{graphicx}
\usepackage{latexsym,amsfonts}
\begin{document}

\title{Neutrino masses from the GSI anomaly}

\author{A. N. Ivanov $^{a,b}$\thanks{E-mail: ivanov@kph.tuwien.ac.at},
      E.  L. Kryshen${^c}$\,\thanks{E--mail: E.Kryshen@gsi.de},
      M. Pitschmann${^a}$\,\thanks{E--mail:
      pitschmann@kph.tuwien.ac.at }, and
      P. Kienle$^{b,d}$\thanks{E-mail: Paul.Kienle@ph.tum.de}, \\
      \addressmark{$^a$Atominstitut der \"Osterreichischen
      Universit\"aten, Technische Universit\"at Wien, Wiedner
      Hauptstra\ss e 8-10, A-1040 Wien, \"Osterreich, \\ $^b$Stefan
      Meyer Institut f\"ur subatomare Physik, \"Osterreichische
      Akademie der Wissenschaften, Boltzmanngasse 3, A-1090, Wien,
      \"Osterreich},\\ $^c$ Petersburg Nuclear Physics Institute,
      188300 Gatchina, Orlova roscha 1, Russian
      Federation,\\$^d$Excellence Cluster Universe Technische
      Universit\"at M\"unchen, D-85748 Garching, Germany\\ }

\date{\today}

\begin{abstract}
   We investigate the influence of the strong Coulomb field of a heavy
 nucleus on massive neutrinos, produced in the K--shell electron
 capture ($EC$) decays of the H--like ${^{140}}{\rm Pr}^{58+}$ and
 ${^{142}}{\rm Pm}^{60+}$ ions. The corrections to the neutrino masses
 due to virtually produced charged lepton $W$--boson pairs in the
 strong Coulomb field of a nucleus with charge $Ze$ are calculated and
 discussed with respect to their influence on the period of the
 time--modulation of the number of daughter ions, observed recently in
 the $EC$--decays of the H--like ${^{140}}{\rm Pr}^{58+}$ and
 ${^{142}}{\rm Pm}^{60+}$ions at GSI in Darmstadt. These corrections
 explain the 2.9 times higher difference of the squared neutrino
 masses obtained from the time--modulation of the $EC$--decays with
 respect to the value deduced from the antineutrino--oscillation
 experiments of KamLAND. The values of neutrino masses are calculated.

PACS: 12.15.Ff, 13.15.+g, 23.40.Bw, 26.65.+t
\end{abstract}

\maketitle

The experimental investigation of the $EC$--decays of the H--like
ions ${^{140}}{\rm Pr}^{58+}$ and ${^{142}}{\rm Pm}^{60+}$, i.e.
${^{140}}{\rm Pr}^{58+} \to {^{140}}{\rm Ce}^{58+} + \nu_e$ and
${^{142}}{\rm Pm}^{60+} \to {^{142}}{\rm Nd}^{60+} + \nu_e$, carried out
in the Experimental Storage Ring (ESR) at GSI in Darmstadt
\cite{GSI2}, showed a modulation in time with periods $T_{EC} \simeq
7\,{\rm s}$ of the rate of the number of daughter ions.  Since the
rate of the number of daughter ions is defined by
\begin{eqnarray}\label{label1}
  \frac{dN^{EC}_d(t)}{dt} = \lambda_{EC}(t)\, N_m(t),
\end{eqnarray}
where $\lambda_{EC}(t)$ is the $EC$--decay rate and $N_m(t)$ is the
number of mother ions ${^{140}}{\rm Pr}^{58+}$ or ${^{142}}{\rm
Pm}^{60+}$, the time--modulation of the rate of $N^{EC}_d(t)$ implies
a periodic time--dependence of the $EC$--decay rate $\lambda_{EC}(t)$.

As has been proposed in \cite{Ivanov2}, such a periodic dependence of
the $EC$--decay rate can be explained by the mass--differences of the
neutrino mass--eigenstates. The period of the time--modulation
$T_{EC}$ has been obtained as
\begin{eqnarray}\label{label2}
T_{EC} =  \frac{4\pi \gamma  M_m}{\Delta m^2_{21}},
\end{eqnarray}
where $M_m$ is the mass of the mother ion, $\gamma = 1.43$ is the
Lorentz factor of the H--like ions \cite{GSI2} and $\Delta m^2_{21} =
m^2_2 - m^2_1$ is the difference of the squared neutrino masses $m_2$
and $m_1$.

For $T_{EC} = 7.06(8)\,{\rm s}$ \cite{GSI2}, measured for the H--like
${^{140}}{\rm Pr}^{58+}$ ion, we have got $(\Delta m^2_{21})_{\rm
GSI} = 2.18(3)\times 10^{-4}\,{\rm eV}^2$ \cite{Ivanov2}, which is by
a factor 2.9 larger than $(\Delta m^2_{21})_{\rm KL} = 7.59(21)\times
10^{-5}\,{\rm eV^2}$ \cite{KL08}, used also for the global analysis of
the solar--neutrino and KamLAND experimental data \cite{PDG06} (see
also \cite{NO3}). For the first time the value $(\Delta m^2_{21})_{\rm
GSI} \simeq 2.25\times 10^{-4}\,{\rm eV}^2$ has been obtained by
Kleinert and Kienle within the neutrino--pulsating vacuum approach
\cite{Kleinert}. The same estimate for $\Delta m^2_{21}$ one can get
by using the period of the time--modulation derived by Lipkin
\cite{Lipkin}.

For the understanding of such a discrepancy we propose the following
mechanism. In the $EC$--decay of a H--like heavy ion a daughter ion
with electric charge $Ze$ and a massive neutrino are produced.  Since
a characteristic energy scale is of order of a few $10^{-15}\,{\rm
eV}$ \cite{Ivanov2}, one possible solution of the discrepancy between
$(\Delta m^2_{21})_{\rm GSI}$ and $(\Delta m^2_{21})_{\rm KL}$ is that a
massive neutrino gets a correction to its mass, caused by its
interaction with the strong Coulomb field of the daughter ion due to
virtually produced $\ell^- W^+$ pairs, where $\ell^- = e^-, \mu^-$ and
$\tau^-$ is a negatively charged lepton and $W^+$--boson, as an
intermediate state. The Feynman diagrams of the process are depicted
in Fig.\,1
\begin{figure}[t]
\centering
\includegraphics[width=\linewidth]{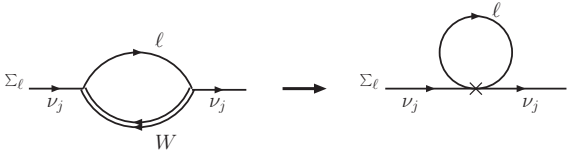}
\caption{Feynman diagrams, defining corrections to the mass of a
massive neutrino in a strong Coulomb field of a nucleus with charge
$Ze$.}
\end{figure}
with the Green functions of virtual charged leptons in the strong
Coulomb field \cite{Coulomb1,Coulomb2}.

For the calculation of the diagrams in Fig.\,1 we use the weak
leptonic interaction \cite{NO2}
\begin{eqnarray}\label{label3}
&&{\cal L}_{W}(x) = -\frac{G_F}{\sqrt{2}}\sum_{j\ell}\sum_{\ell'
j'}U_{j\ell}U^*_{\ell' j'}\nonumber\\
&&\times\,[\bar{\psi}_{\nu_j}(x)\gamma^{\mu}(1 -
\gamma^5)\psi_{\ell}(x)]\nonumber\\
&&\times\,[\bar{\psi}_{\ell'}(x)\gamma_{\mu}(1 -
\gamma^5)\psi_{\nu_{j'}}(x)],
\end{eqnarray}
defined by the W--boson exchange, where $x = (t,\vec{r}\,)$, $G_F$ is
the Fermi constant, $\psi_{\nu_j}(x)$ and $\psi_{\ell}(x)$ are
operators of the neutrino $\nu_j$ and lepton fields $\ell = e^-$,
$\mu^-$ and $\tau^-$, respectively, and $U_{\ell j}$ are the elements
of the unitary neutrino--flavour mixing matrix $U$ \cite{PDG06}. In
our analysis neutrinos $\nu_j\,(j = 1,2,3)$ are Dirac particles with
masses $m_j\,(j = 1,2,3)$, respectively \cite{PDG06}.

A correction $\delta m_j$ to the neutrino mass, induced by the
interaction of the neutrino $\nu_j$ with a strong Coulomb field of a
nucleus, is defined by
\begin{eqnarray}\label{label4}
\delta m_j(r) = \sum_{\ell} U_{j\ell}U^*_{\ell
j}\,{\cal M}_{\ell}(r),
\end{eqnarray}
where we have denoted
\begin{eqnarray}\label{label5}
{\cal M}_{\ell}(r) = i\,\sqrt{2}\,G_F\int^{+i\infty}_{-i\infty}
\frac{d E}{2\pi }{\rm tr}\{
G_{\ell}(\vec{r},\vec{r}; E)\gamma^0\}.
\end{eqnarray}
Here $G_{\ell}(\vec{r},\vec{r}; E)$ is the energy--dependent Green
function of the negatively charged leptons $\ell^-$ in a strong Coulomb
field, produced by a positive electric charge $Ze$ \cite{Coulomb1,Coulomb2}.

Using the results, obtained in \cite{Coulomb2}, we get
\begin{eqnarray}\label{label6}
 \hspace{-0.3in} &&{\cal M}_{\ell}(r) =
  \sqrt{2}\,G_F\,\frac{m_{\ell}}{\pi^2 r^2} \sum^{\infty}_{n = 1}n
  \int^{\infty}_0 \int^{\infty}_0dx\,dt\nonumber\\
\hspace{-0.3in}&&\times\,e^{- 2 m_{\ell}r\sqrt{x^2
  + 1}\coth t}\,\Big\{2 Z\alpha \coth t \cos\Big(\frac{2 Z\alpha x t}{\sqrt{x^2 +
  1}}\Big)\nonumber\\
   \hspace{-0.3in}&&\times\,\tilde{I}_{2\nu}\Big(\frac{2m_{\ell}r \sqrt{x^2
  + 1}}{\sinh t}\Big) - \sin\Big(\frac{2 Z\alpha x t}{\sqrt{x^2 +
  1}}\Big)\nonumber\\ \hspace{-0.3in}&&\times\,\Big[\frac{2m_{\ell}r
  x}{\sinh t}\tilde{I}_{2\nu + 1}\Big(\frac{2m_{\ell}r \sqrt{x^2 + 1}}{\sinh
  t}\Big) + \frac{2 \nu x}{\sqrt{x^2 + 1}}\nonumber\\ \hspace{-0.3in}
  &&\times\,\tilde{I}_{2\nu }\Big(\frac{2m_{\ell}r \sqrt{x^2 + 1}}{\sinh
  t}\Big)\Big]\Big\},
\end{eqnarray}
where $\nu = \sqrt{n^2 -(Z\alpha)^2}$ and $I_{\mu}(z)$ is a modified
Bessel function \cite{HB}, $\tilde{I}_{2\nu + 1}(z) = I_{2\nu + 1}(z)
- I_{2n +1 }(z)$ and $\tilde{I}_{2\nu}(z) = I_{2\nu}(z) - I_{2n}(z)$.
We would like to notice that at $Z\alpha \to 0$ the corrections to the
neutrino masses vanish as ${\cal M}_{\ell}(r) \to 0$. Hence, a
non--vanishing correction to the massive neutrino mass appears only
due to the Coulomb field.  Since at $r \to \infty$ the corrections
introduced by Eq.(\ref{label5}) vanish rapidly, so that in the
subsequent interactions \cite{Ivanov3} the neutrino $\nu_j$ should be
with a proper mass $m_j$. The very rapid vanishing (see Fig.\,2)
\begin{figure}[t]
\centering
\includegraphics[width=\linewidth]{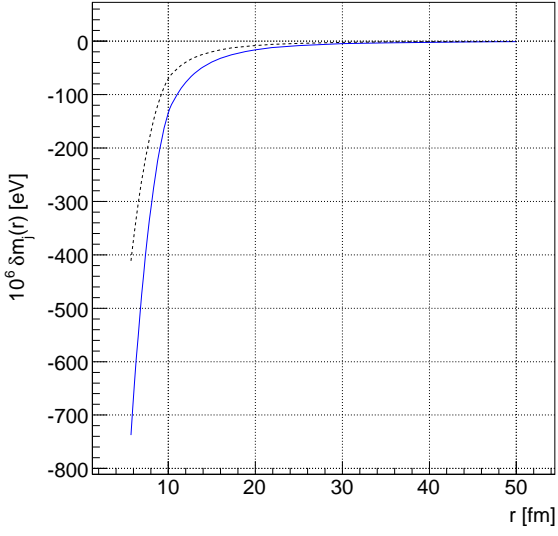}
\caption{The corrections to the neutrino masses, caused by a strong
nuclear Coulomb field, where $\delta m_1(r)$ and $\delta m_2(r)$ are
presented by the solid and dotted line, respectively.}
\end{figure} of
the $\delta m_j(r)$ with $r$ makes it reasonable to take into account
the influence of the correction only at the nuclear surface
\cite{BS67} just after the production of the massive neutrino $\nu_j$
and the daughter ion\,\footnote{The amplitude of the $EC$--decay is
proportional to \cite{Ivanov2,Ivanov1}
\begin{eqnarray*}
\hspace{-0.3in}&&M(m \to d\,\nu_j)(t) \propto \int
d^3x\,\Psi^*_d(r)\Psi_m(r) \psi^{(Z)}_{1s}(r) e^{ i E_{\nu_j}(r) t} =\nonumber\\
\hspace{-0.3in}&&= e^{ i E_{\nu_j}(R) t} \langle
\psi^{(Z)}_{1s}\rangle\, {\cal M}_{\rm GT},
\end{eqnarray*}
where $E_{\nu_j}(r) = \sqrt{\vec{k}^{\,2}_j + (m_j + \delta
m_j(r))^2}$. Using an analogy between the Fermi--Dirac distribution
function and the Woods--Saxon shape of the nuclear density \cite{GT}
and following \cite{LL87} one can show that the relation $\langle
\psi^{(Z)}_{1s}\,e^{\,iE_{\nu_j} t}\rangle = \langle
\psi^{(Z)}_{1s}\rangle\,e^{\,i\,E_{\nu_j}(R) t}$ is fulfilled with an
accuracy better than $1\,\%$. For the confirmation of the validity of
this relation we refer also on \cite{BS67}.}.

At the nuclear radius $r = R = 5.712\,{\rm fm}$ \cite{Ivanov1}, we get
\begin{eqnarray}\label{label7}
{\cal M}_{e^-}(R) &=&-\,2.02\times 10^{-3}\,{\rm eV},\nonumber\\
{\cal M}_{\mu^-}(R)&=&-\,5.16\times 10^{-4}\,{\rm eV},\nonumber\\
{\cal M}_{\tau^-}(R)&=&-\,3.88\times 10^{-5}\,{\rm eV}
\end{eqnarray}
 with an electron $e^-$ , muon $\mu^-$ and $\tau^-$--lepton in the
intermediate state, respectively.  The corrections to the neutrino
masses are equal to
\begin{eqnarray}\label{label8}
\delta m_1(r) &=& \cos^2\theta_{12}{\cal M}_{e^-}(r) +
\sin^2\theta_{12}(\cos^2\theta_{23}\nonumber\\
&&\times\,{\cal M}_{\mu^-}(r) +
\sin^2\theta_{23}{\cal M}_{\tau^-}(r)),\nonumber\\ 
\delta m_2(r) &=&
\sin^2\theta_{12}{\cal M}_{e^-}(r) +
\cos^2\theta_{12}(\cos^2\theta_{23}\nonumber\\
&&\times\,{\cal M}_{\mu^-}(r) +
\sin^2\theta_{23}{\cal M}_{\tau^-}(r)),
\end{eqnarray}
where $\theta_{12}$ and $\theta_{23}$ are mixing angles. The
corrections to the neutrino masses Eq.(\ref{label8}) are defined for
$\theta_{13} = 0$ \cite{Ivanov2} (see also \cite{NO3}). Then, setting
$\theta_{12} = 34^0$ and $\theta_{23} = 45^0$ \cite{NO3} we obtain
\begin{eqnarray}\label{label9}
\delta m_1(R) &=& -\,14.74\times 10^{-4}\,{\rm eV},\nonumber\\ \delta
m_2(R) &=& -\,8.22\times 10^{-4}\,{\rm eV}.
\end{eqnarray}
The period of modulation is thus redefined as 
\begin{eqnarray}\label{label10}
T_{EC} = \frac{4\pi \gamma M_m}{(m_2 + \delta m_2(R))^2 - (m_1 + \delta m_1(R))^2 }.
\end{eqnarray}
Neglecting the contributions of $(\delta m_j(R))^2$ we transcribe the
denominator into the form
\begin{eqnarray}\label{label11}
\hspace{-0.3in}&&\delta m^2_2(R) - \delta m^2_1(R) =  (\Delta
m^2_{21})_{\rm GSI} - (\Delta m^2_{21})_{\rm KL}\nonumber\\
\hspace{-0.3in}&&\hspace{0.3in}+ (\delta m_1(R))^2 - (\delta m_2(R))^2,
\end{eqnarray}
where $\delta m^2_j(R) = 2 m_j\delta m_j(R)$.  Using the numerical
values of the corrections Eq.(\ref{label9}), $(\Delta m^2_{21})_{\rm
GSI} = 2.20\times 10^{-4}\,{\rm eV}^2$, $(\Delta m^2_{21})_{\rm KL} =
7.59\times 10^{-5}\,{\rm eV}^2$ and a relation $m_2 - m_1 = (\Delta
m^2_{21})_{\rm KL}/(m_2 + m_1)$ we solve Eq.(\ref{label11}) and get the
following values for neutrino masses
\begin{eqnarray}\label{label12}
m_2 &=& 0.11 + 0.82\times 10^{-4}\,{\rm eV},\nonumber\\
m_1 &=& 0.11 + 4.26\times 10^{-4}\,{\rm eV}.
\end{eqnarray}
The mass $m_3$ of the neutrino $\nu_3$ is
\begin{eqnarray}\label{label13}
m_3 = 0.12 + 8.05\times 10^{-4}\,{\rm eV}.
\end{eqnarray}
We obtain it using Eq.(\ref{label12}) and the experimental value
$\Delta m^2_{32} = 2.4\times 10^{-3}\,{\rm eV}^2$ \cite{NO2}. The sum
of neutrino masses amounts to
\begin{eqnarray}\label{label14}
\sum_{j = 1,2,3}m_j = 0.34\,{\rm eV},
\end{eqnarray}
which agrees well with the upper limit $\sum_jm_j < 1\,{\rm eV}$
\cite{PDG06}.

We have shown that an interaction of virtually produced $\ell^- W^+$
pairs $\nu_j \to \sum_{\ell} U_{j \ell}\ell^- W^+$ of massive
neutrinos $\nu_j$ in the strong Coulomb field of the daughter ion can
induce certain corrections to neutrino masses, which allow to
reconcile the value $(\Delta m^2_{21})_{\rm GSI} = 2.18(3)\times
10^{-4}\,{\rm eV^2 }$ \cite{Ivanov2}, deduced from the period of the
time--modulation of the rate of the number of daughter ions in the
$EC$--decays of the H--like ions ${^{140}}{\rm Pr}^{58+}$ and
${^{142}}{\rm Pm}^{60+}$, with $(\Delta m^2_{21})_{\rm KL} =
7.59(21)\times 10^{-5}\,{\rm eV}^2$ \cite{PDG06,NO3}, obtained as a
best--fit of the global analysis of the solar--neutrino and KamLAND
experimental data \cite{PDG06} (see also \cite{NO3}). We would like to
notice that for the calculation of the corrections to neutrino masses
we have taken into account the contribution of the $W^+$--boson
exchange only. The contribution of the $Z$--boson exchange is
proportional to the constant $g_V = -\,0.040\pm 0.015$
\cite{PDG06}. This means that the corrections to neutrino masses,
caused by the $Z$--boson exchanges, are smaller compared with
corrections, which can be caused by the experimental uncertainties of
the mixing angles $\theta_{12} = 33.9^{+2.4}_{-2.2}\,{\rm degrees}$
and $\theta_{23} \le 45\,{\rm degrees}$ \cite{PDG06}.

The proposed change of the neutrino masses together with the
experimental data on the time--modulation of the rate of the number of
daughter ions in the $EC$--decays of the H--like ions and $(\Delta
m^2_{21})_{\rm KL} = 7.59\times 10^{-5}\,{\rm eV}^2$ allows to
estimate the values of neutrino masses $m_j \simeq 0.11\,{\rm eV}$
agreeing well with the constraint on the sum of neutrino masses $\sum
m_j < 1\,{\rm eV}$ \cite{PDG06}. The value of the heaviest neutrino
mass $m_3 = 0.12 + 8.05 \times 10^{-4}\,{\rm eV}$ satisfies also the
constraint $0.04 < m_3 < 0.40\,{\rm eV}$ \cite{PDG06}.


\begin{thebibliography}{9}
\bibitem{GT} ${\cal M}_{\rm GT}$ is the nuclear matrix element of the
Gamow--Teller transition defined by ${\cal M}_{\rm GT} = - 2 g_A G_F
V_{ud}\int d^3x \Psi^*_d(r)\Psi_m(r)$, where $\Psi^*_d(r)$ and
$\Psi_m(r)$ are the wave functions of the daughter and mother nuclei,
respectively.  We set $\Psi^*_d(r)\Psi_m(r) \sim \rho(r)$, where
$\rho(r)$ has the Woods--Saxon shape with a nuclear radius $R =
1.1\,A^{1/3}\,{\rm fm}$ and a parameter $a = 0.50\,{\rm fm}$,
characterising the diffuseness (or the width) of the nuclear surface.
\end{thebibliography}
\end{document}